\documentstyle[epsfig,12pt]{article}

\begin{document}

\hoffset = -1truecm \voffset = -2truecm \baselineskip = 10 mm

\title{\bf Particle multiplicities at LHC and deviations from limiting fragmentation}

\author{{\bf Jianhong Ruan} and {\bf Wei Zhu}\\
\normalsize Department of Physics, East China Normal University,
Shanghai 200062, P.R. China \\
}

\date{}

\newpage

\maketitle

\vskip 3truecm

\begin{abstract}

    The pseudorapidity density of charged particles produced
at LHC collisions are predicted by using two complementary
production mechanisms with a set of consistent integrated and
unintegrated parton distributions. We discuss the limiting
fragmentation hypothesis and its possible violation, and we compare
our model with other partonic models.

\end{abstract}

PACS number(s): 12.38.Bx, 13.60.Hb, 24.85.+p, 25.75.-q

\newpage
\begin{center}
\section{Introduction}
\end{center}

     Particle multiplicity distribution is one of the first
measurements to be taken at the CERN Large Hadron Collider (LHC).
The upcoming data on its energy, centrality, and rapidity dependence
are expected to discriminate various (integrated and unintegrated)
parton distributions, which are basic qualities analyzing
high-energy reactions on the parton level. The Bjorken variable $x$
of gluons may reach very small values at LHC energies. Therefore,
the nonlinear corrections of the initial gluon correlations to the
QCD evolution equations should be considered in any available parton
distributions of the LHC physics [1].

    One of the striking predictions of nonlinear QCD evolution equations
is the saturation solution of the
Jalilian-Marian-Iancu-McLerran-Weigert-Leonidov-Kovner (JIMWLK)
equation [2], where the unintegrated gluon distribution is
absolutely flat in $k_t$-space at $k_t<Q_s$; $Q_s$ is the saturation
scale and the relating phenomenon is called the color glass
condensate (CGC). The above saturation solution was proved in the
numerical solutions of the JIMWLK equation [3]. Instead of the
complicated JIMWLK equation, which is equivalent to an infinite,
coupled hierarchy of evolution equations (the Balitsky hierarchies),
much of the physics content has been studied with the help of
approximations, such as the Balitsky-Kovchegov (BK) equation [4], or
outright models that capture the main features as they affect
specific phenomena, such as the Kharzeev-Levin (KL) model [5] does
for the present context.

    The KL model mimics a possible saturation solution of the JIMWK equation, and
it assumes that the scale $Q_s^2(x)\equiv Q^2_0(x_0/x)^{\lambda}$.
Under some additional assumptions, Kharzeev, Levin, and Nardi (KLN)
use this model successfully to explain some of the climactical data
probed at the BNL Relativistic Heavy-Ion Collider (RHIC) and to
predict LHC physics [6]. The BK equation is regarded as the leading
order approximation of the JIMWLK equation and preserves the
saturation and scaling features. Albacete uses the BK equation to
predict the pseudorapidity density of charged particles produced in
central $Pb-Pb$ collisions at LHC [7].

       However, both the KL model and BK equation work at a very small $x$ range,
(for say, $x<x_0,~ x_0\ll 10^{-1}$), where the saturation begins
work. The $x$-dependence of the gluon density at $x>x_0$ in these
works [6,7] are simply fixed to be $(1-x)^4$, which is un-evolution.
As we have pointed out, the parton distributions at intermediate and
lager $x$ influence the shape of the rapidity spectrum [8]. In fact,
Szczurek uses the same KL model but with different factors $\sim
(1-x)^{5-7}$ and obtains a narrower pseudo-rapidity distribution
[9]. Anyway, a fixed form $\sim(1-x)^4$ for an available gluon
distribution is too rough. Obviously, as a complement to the
above-mentioned saturation models, it is necessary to consider the
parton distributions, which are well defined in a broad kinematic
range.

       Besides, some works [6,7] have used a single production mechanism,
i.e., gluon-gluon fusion $gg\rightarrow g$, which is proposed by
Gribov, Levin and Ryskin (GLR) [10] to predict particle production.
However, the single-particle inclusive spectrum shows that its
rapidity distribution has three distinct regions: a central region
with two (project and target) fragmentation regions [11]. The GLR
model is expected to dominate the processes at the central region.
Therefore, other collision dynamics in the fragmentation regions
should be considered. There are several two-component models for
hadron collisions. For example, the HIJING model [12] is such a
two-component model. This model uses the parameterized integrated
parton distributions to compute multiple minijet production and
incorporates the Lund string model [13] to model soft beam jet
fragmentation. However, the production of minijets at the central
region should be described by the unintegrated gluon distribution
rather than the integrated gluon density [9]. Moreover, the string
model is irrelevant to the parton distributions. In the same model,
the hard and soft components based on different basic physical
parameters may lose some interesting information about heavy-ion
collisions.

    In this work, we use a two-component model and a set of
consistent parton distributions to improve the above-mentioned
situation of the present models. Basically, we use two complementary
production mechanisms: the hard gluon-gluon fusion [10] and the soft
quark recombination [14]. Our picture is as follows. At sufficiently
high-energy hadron-hadron collisions, particles produced in hard
gluon-gluon fusions are distributed in a region around midrapidity,
and these initial gluons are described by the unintegrated gluon
distribution in both colliding hadrons. On the other hand, the
valence quarks tend to fly through with their original integrated
distributions and they are hadronized by recombining with additional
low-$x$ sea quarks from the central region. The resulting soft
recombined particles dominate the fragmentation region. We shall
present the related formula of two production mechanisms in Sec.II.
Because gluon fusion and quark recombination use unintegrated and
integrated parton distributions, respectively, a set of these parton
distributions, which are defined in a broad kinematic range are
necessary. In our previous work [8], we proposed such distributions
in protons and heavy nuclei by using a modified Dokshitzer-Gribov-
Lipatov-Altarelli-Parisi (MD-DGLAP) equation [15], which
incorporates the shadowing and antishadowing corrections. Unlike the
JIMWLK and BK equations, MD-DGLAP equation works in a broad
pre-saturation range. We shall use these parton distributions to
explain the particle multiplicity distributions in UA5 and RHIC
data, then we will predict the particle multiplicity distributions
at LHC in Secs. III and IV. Unfortunately, we can not theoretically
predict the energy dependence of the normalization of the hadronic
distributions, which is fixed using the entirely phenomenological
parametrization of existing data in this work. This simple method
may need modification after we obtain the LHC data at midrapidity.

    In Sec. V we discuss some interesting properties of rapidity
distributions. Limiting fragmentation [16] and rapidity plateau [17]
are characteristics of two components of particle production. We
find that the limiting fragmentation hypothesis, which generally
appears in present data of hadron collisions is partly violated if
the observations are over a wide range between the RHIC and LHC
energies. An explanation of limiting fragmentation and its violation
in partonic picture is given. On the other hand, we propose that a
possible quark gluon plasma (QGP) effect may deform the shape of the
central plateau. We also discuss the nuclear shadowing effects in
heavy-ion collisions. The comparisons of our predictions with those
of saturation models are given in this section. The above-mentioned
properties of particle multiplicity distributions in the partonic
models of hadron collisions, except for their height at midrapidity,
are dominated by the parton distributions. Therefore, the results
potentially tell us how parton distributions evolve at high
energies. The last section is a short summary of this study.

\newpage
\begin{center}
\section{Two-component model}
\end{center}

    The single-particle inclusive spectrum shows a rapidity
distribution with three distinct regions: a central region and two
(project and target) fragmentation regions. We assume that the
hadrons produced in the central region (small $x$ and large $k_t$)
are produced from the hadronization of the gluons in the
$gg\rightarrow g$ mechanism [10], while the particles in the
fragmentation region are formed by the valence quarks according to
the quark recombination model [14]. The related formulas are
summarized as follows.

    Component I. The cross section for inclusive gluon production in
$pp\rightarrow g$ through the gluonic mechanism $gg\rightarrow g$ at
sufficiently high energy reads [10]

$$\frac{d\sigma_{p-p}^{I}(y,p_{t,g})}{dyd^2p_{t,g}}$$
$$=\frac{4N_c}{N_c^2-1}\frac{1}{p_{t,g}^2}
\int d^2q_{t,g}\alpha_s(\Omega)
F_g^p\left(x_1,\left(\frac{p_{t,g}+q_{t,g}}{2}\right)^2,p^2_{t,g}\right)
F_g^p\left(x_2,\left(\frac{p_{t,g}-q_{t,g}}{2}\right)^2,p^2_{t,g}\right),\eqno(1)$$
where $\Omega=\max(k^2_{1t},k^2_{2t},p^2_{t,g})$ ,
$k^2_{1,t}=\frac{1}{4}(p_{t,g}+q_{t,g})^2$ and
$k^2_{2,t}=\frac{1}{4}(p_{t,g}-q_{t,g})^2$; The rapidity $y$ of the
produced gluon in the center-of-mass frame of $p-p$ collisions is
defined by

$$x_{1/2}=\frac{p_{t,g}}{\sqrt{s}}\cdot\exp(\pm y);\eqno(2)$$ $F_g^p(x,k_t^2,p_t^2)$
is the two scale unintegrated gluon distribution in the proton. A
general relation between integrated and unintegrated parton
distributions is

   $$\int^{\mu^2}_0dk^2_tF_a^p(x,k^2_t,\mu^2)
=xa_p(x,\mu^2), \eqno(3)$$ where $a_p(x,\mu^2)=v_p(x,\mu^2)$,
$s_p(x,\mu^2)$ and $g_p(x,\mu^2)$ imply the integrated valence
quark, sea quark and gluon distributions in the proton.

    In experiments a good identification of particles is to measure pseudorapidity.
The relation between the rapidity $y$ and pseudorapidity $\eta$ for
massive particles is

$$y=\frac{1}{2}\ln \left[\frac{\sqrt{\frac{m^2_{eff}+p^2_t}{p^2_t}+\sinh^2 \eta}+\sinh \eta}
{\sqrt{\frac{m^2_{eff}+p^2_t}{p^2_t}+\sinh^2\eta}-\sinh \eta}
\right], \eqno(4)$$ where $m_{eff}$ is the typical invariant mass of
the gluon mini-jet.

    For avoiding the complicate hadronization dynamics, similar to Ref. [9], we
use local parton-hadron duality, i.e., the rapidity distribution of
particles is identical to the rapidity distribution of gluons:
$\eta_g=\eta_h\equiv \eta.$ Thus, the pseudorapidity density of
produced charged particles in $p-p$ collisions is given by

$$\frac{dN_{p-p}^{I}}{d\eta }$$
$$=\frac{1}{\sigma_{in}}\int d^2p_{t,h}\frac{d \sigma^{I}_{p-p}(\eta, p_{t,h})}{d\eta
d^2p_{t,h}}$$
$$=\frac{1}{\sigma_{in}}\int\frac{dz}{z}d^2p_{t,h}J_1(\eta_g;p_{t,h};m_{eff})
\left.D(z,p_{t,h})\delta^2(p_{t,h}-zp_{t,g})\frac{d
\sigma_{p-p}^{I}(y,p_{t,g})}{dy
d^2p_{t,g}}\right|_{y\rightarrow\eta_g}$$
$$\equiv c(\sqrt s)\int d^2p_{t,g}J_1(\eta;p_{t,g};m_{eff})
\left.\frac{d \sigma_{p-p}^{I}(y,p_{t,g})}{dy
d^2p_{t,g}}\right|_{y\rightarrow\eta}, \eqno(5)$$ where the Jacobian
is

$$J_1(\eta;p_t;m_{eff})=\frac{\cosh\eta}{\sqrt{\frac{m^2_{eff}+p^2_t}{p^2_t}+\sinh^2\eta}},\eqno(6) $$
and we neglect the fragmentation function $D(z,p_{t,h})$.
Corrections to the kinematics due to the hadron mass are considered
by replacing $p^2_t\rightarrow p^2_t+m^2_{eff}$ in the evaluation of
$x_{1/2}$. Assuming pions in $p-p$ collisions are produced via
$\rho$-resonance, we take $m_{eff}=770 MeV$.

    Component II. According to the quark recombination model [14], the valence
quarks of incident proton tend to fly through the central region
with their original momentum fraction. These valence quarks
recombine with lower $p_t$ antiquarks and produce the outgoing
hadrons in the fragmentation region. The quark recombination model
has explained successfully the meson inclusive distributions and the
leading particle effects in the fragmentation region below RHIC
energies [18].

     The cross section for inclusive pion production in
$p-p$ collisions in the quark recombination model is

$$\frac{1}{\sigma_{in}}\frac{d\sigma_{p-p}^{II}}{dxdp_t^2}$$
$$=6\frac{1-x}{x}\int^{x}_0dx_1x_1v_p(x_1,p_t^2)
\frac{1}{2}(1+\delta)(x-x_1)s_p(x-x_1,p_t^2),\eqno(7)$$ where
$\delta s_p(x,p^2_t)$ is the distribution of additional sea quarks
in the central region and we assume that it has the form like
$s_p(x,p^2_t)$.

    We introduce the rapidity
for pions in the recombination processes as

$$y=\ln x-\ln\frac{\sqrt{m^2_{\pi}+p_t^2}}{\sqrt{s}}.\eqno(8)$$
Thus, we have

$$\frac{1}{\sigma_{in}}\frac{d\sigma_{p-p}^{II}(y,p_t)}{dydp^2_t}=\left.
J_{II}(y;p_t;m_{\pi})\frac{1}{\sigma_{in}}\frac{d\sigma_{p-p}^{II}(x,p_t)}{dxdp^2_t}\right|_{x\rightarrow
y},\eqno(9)$$ with a new Jacobian

$$J_{II}(y;p_t;m_{\pi})=\frac{\partial x}{\partial
y}=\frac{\sqrt{p^2_t+m^2_{\pi}}}{\sqrt {s}}e^{y},\eqno(10)$$ and
they lead to

$$\frac{dN_{p-p}^{II}}{d\eta }=\frac{1}
{\sigma_{in}}\int dp^2_t\frac{d \sigma_{p-p}^{II}(\eta,p_t)}{d\eta
dp^2_t }= \frac{1}{\sigma_{in}}\int dp^2_t
J_{II}(\eta;p_t;m_{\pi})\left.\frac{d
\sigma_{p-p}^{II}(y,p_t)}{dydp^2_t }\right|_{y\rightarrow\eta
},\eqno(11)$$ where we integrate over transverse momenta in Eq. (11)
at $p_t<1 GeV$, since the recombination model works at lower $p_t$
range.

    Summing the contributions of two components, we have the total
distribution

$$\frac{dN_{p-p}}{d\eta }=\frac{dN_{p-p}^{I}}{d\eta }
+\frac{dN_{p-p}^{II}}{d\eta }.\eqno(12)$$

    The production mechanisms (I) and (II) use
unintegrated and integrated parton distributions, respectively. In
particularly, the gluon momentum fraction in Eqs. (1) and (7)
contain both smaller and larger $x$ regions. Therefore, a set of
consistent integrated and unintegrated parton distributions in
protons and heavy nuclei, which are defined in a broad kinematic
region, are necessary. Fortunately, such parton distributions are
proposed in Ref. [8], where the integrated parton distributions are
evolved by using a modified DGLAP equation [15] in a whole
pre-saturation region; while the unintegrated parton distributions
are obtained directly from these integrated parton distributions
using the Kimber, Martin and Ryskin (KMR) scheme [19]. We shall use
these parton distributions to predict the particle multiplicity
distributions.

\newpage
\begin{center}
\section{Proton-proton collisions}
\end{center}

    We calculate the pion distributions in $p-p$ collisions
using the two component model and compare the results with the data.
We need the values of the total inelastic cross section
$\sigma_{in}$, which is included by the coefficient $c(\sqrt s)$ of
Eq. (5). This is rather complicated at the parton level since it
contains the nonperturbative information. In this work, we use the
midrapidity density for $p-p(\overline{p})$ collisions to estimate
the values of $c(\sqrt s)$. The former has been parameterized by the
UA5 [20] and CDF [21] collaborations as

$$\left.\frac{dN_{p-p}}{d\eta
}\right|_{\eta=0}=2.5-0.25\ln s+0.023\ln^2s. \eqno(13)$$ This purely
empirical parametrization is fitted in a broad range from
$\sqrt{s}=15$GeV to $1.8$TeV and we extend it to the LHC energies.
The resulting $c(\sqrt s)$ is shown in Fig.1.

    Figure 2 shows our results for $dN_{p-p}/d\eta$ bellow
LHC energies with $m_{eff}=770$MeV (solid curves). The data are
taken from Ref.[20]. For comparison, we also draw the distributions
with $m_{eff}=0$. We find that the shape of the central rapidity
plateau relates sensitively to the value of parameter $m_{eff}$. As
shown in Fig.2, corresponding to $m_{eff}=770$MeV, the full rapidity
plateau has two peaks. During the reduction of the $m_{eff}$ value,
the rapidity plateau is flattened at $\vert\eta\vert<2$ and even
disappears.

    To illustrate the contributions from two production
components in our model, in Figs.3 and 4 we use dashed and dotted
curves to indicate the contributions from the gluon fusion and quark
recombination models, respectively. Although the saturation models
[6,7] have used a single gluon fusion mechanism to reproduce these
data, they both take a fixed gluon distribution $\sim(1-x)^4$ at the
pre-saturation range. Now we use a reasonable gluon distribution
instead of $(1-x)^4$. One can find that the resulting dashed curves
are narrower than the solid curves (which are consistent with the
data in Fig. 2) in Figs. 3 and 4, and this implies that an
additional contribution from the fragmentation regions is necessary.

    We predict the pion distributions in $p-p$ collisions at LHC
energies in Figs.5-9. Although the contributions of the quark
recombination model (dotted curves) are generally smaller than that
of the gluon-gluon fusion model (dashed curves), the contributions
of the quark recombination still can not be neglected in the
fragmentation region.

    Several partonic models have predicted the pion distributions in $p-p$
collisions at LHC energies. It is interest to compare our results
with them. Figure 10 presents the predictions of KLN work [6]
(dashed curve) and the comparison with our result (solid curve).
Figure 11 compares the comparisons of our prediction with those of
the PYTHIA model, which is based on string fragmentation mechanism
[22], and PHOJET model, which uses a Pomeron exchange [23]. Figure
12 is a similar comparison with the ultrarelativistic quantum
molecular dynamics (UrQMD) model [24]. We think that the comparisons
with different models can provide useful knowledge about the
unintegrated gluon distribution in the proton and a correct picture
of hadron collisions.

\newpage
\begin{center}
\section{Nucleus-nucleus collisions}
\end{center}

    The nucleus-nucleus collisions are much more complicated than
$p-p$ collisions. The high multiplicities in heavy-ion collisions
typically arise from the large number of nucleon-nucleon collisions.
In the analysis of heavy-ion collision data at highly relativistic
energies, two parameters which characterize the influence of nuclear
geometry are used [25]: (1) the number of participating nucleons
$N_{part}$, which depends on the collision geometry, and (2) the
number of binary nucleon-nucleon collisions $N_{coll}$, or the
average struck number of each participating nucleon as it passes
through the oncoming nucleus,
$\overline{\nu}=N_{coll}/(0.5N_{part})$. In nuclear collisions, the
soft (or hard) component is proportional to the number of
participants $N_{part}$ (or the number of binary collisions
$N_{coll}$) [26]. Essentially, we write

$$\frac{2}{<N_{part}>}\frac{dN_{A-A}}{d\eta}=<\overline{\nu}>
c(\sqrt{s})\int d^2p_t\frac{d\sigma_{A-A}^I(\eta,p_t)}{d\eta
dp_t^2}+ \frac{1}{\sigma_{in}}\int d^2p_t\frac{d
\sigma_{A-A}^{II}(\eta,p_t)}{d\eta dp^2_t }, \eqno(14)$$ where
$<...>$ is an average value in a giving central cut and we only
consider $0-6\%$ cut in this work; $d\sigma_{A-A}$ implies that the
parton distributions are the nuclear parton distributions. According
to the geometric approach in Ref.[27], we take the mean number of
participants $N_{part}=339$ in $Au-Au$ collisions at $\sqrt s=130$
and 200GeV, and $N_{part}=369$ in $Pb-Pb$ collisions at $\sqrt
s=5.5$TeV.

    The value of $\overline{\nu}$ contain the knowledge about the
interaction between two collided nuclei. Glauber modeling in
high-energy nuclear collisions [28] has pointed out that the number
of collisions roughly is

   $$<N_{coll}>\propto <N^{4/3}_{part}>\equiv\lambda <N^{4/3}_{part}>.\eqno(15)$$
On the other hand, the RHIC data [29] present a slow increase of the
coefficient $\lambda$ with energy $\sqrt s$. Using these data about
$dN_{Au-Au}/d\eta\vert_{\eta=0}$ at $\sqrt s=130-200$GeV, we take a
best fitting: $\lambda=29\ln \sqrt{s}$.

   Now we can "predict" the whole distributions in $Au-Au$
collisions at $\sqrt s=130$ and $200$GeV. The results are shown in
Figs.13 and 14. Where we keep temporarily the value of parameter
$m_{eff}=770$MeV. The dashed and dotted curves are the contributions
of the gluon-gluon fusion model and quark recombination model,
respectively. There is a small deviation from the data at $\vert
\eta\vert>5$ since the Fermi motion contributions [30] are neglected
in our considerations.

    Considering the experimental errors, the inconsistency between the
theoretical curves and the RHIC data is still visible, and it
suggests that some factors failed in our above considerations. In
Figs. 15 and 16, we reduce the value of $m_{eff}$. A best fit (solid
curves) requests the parameters $m_{eff}=400$MeV. Compared with our
previous result of $m_{eff}=770$MeV for $p-p$ collisions in Sec.
III, we find that the reduction of $m_{eff}$ at RHIC $Au-Au$
collisions is possible.

    Finally, we calculate the pseudorapidity distributions in $Pb-Pb$
collisions at $\sqrt s=5.5$TeV. The differences between lead and
gold are neglected. Our results are shown by solid curve in Fig.17,
where we take $m_{eff}=400$GeV; the dashed, and dotted curves
correspond to the contributions from gluon-gluon fusion and quark
recombination, respectively.

\newpage
\begin{center}
\section{Discussions}
\end{center}

    (i) Limiting fragmentation, is it universal or not?

        To separate the trivial kinematic broadening of the $dN/d\eta$
distribution from more interesting dynamics, the collision data at
different energies are viewed in the rest frame of one of the
colliding target. Such distributions lead to a striking universality
of multiparticle production--limiting fragmentation. This hypothesis
states that, at high enough collision energy, when effectively
viewed in the target rest frame, $dN/d\eta'$ exhibits longitudinal
scaling and becomes independent of energy in a region around
$\eta'\sim 0$, where $\eta'=\eta-y_{beam}$,
[$y_{beam}=\ln(\sqrt{s}/m_N)$]. The hypothesis of limiting
fragmentation in high energy hadron-hadron collisions was first
suggested in Ref. [16]. From a phenomenological view, the projectile
hadron, when seen in the frame of the target, is Lorentz-contracted
into a very narrow strongly-interacting pancake which passes through
the target, assuming that the total hadronic cross sections would
become constant at large center-of-mass energy. If this occurred,
the excitation and break-up of a hadron would be independent of the
center-of-mass energy and distributions in the fragmentation region
would approach a limiting curve.  We know that the total hadronic
cross-sections are not constant at high energies, therefore,
limiting fragmentation should fail. However, limiting fragmentation
has been observed in a wider region, even extending nearly to
midrapidity, and it is referred to as extended longitudinal scaling
[25].

    From the partonic point of view, longitudinal scaling in
hadron collisions relates to Bjorken scaling of the parton
distributions and the production dynamics. An interesting question
is whether two component models can keep the limiting fragmentation
curve. To answer this question, we plot the shifted pseudorapidity
distributions in central $Au-Au$ collisions at $\sqrt s=130GeV$ and
$200$GeV in Fig.18. The distributions are scaled by $N_{part}/2$ to
remove the effect of the different number of nucleons participating
in the collisions. We find longitudinal scaling (energy
independence) over more than three units of rapidity, extending
nearly to midrapidity and it is consistent with the RHIC data [31].
In Fig. 19, we present the contributions only from the gluon fusion
mechanism, where limiting fragmentation still holds at $\eta'>-1$.

    However, we compare the similar distributions including $Pb-Pb$
collisions at $\sqrt s=5.5$TeV (dotted curve)in Fig. 20. We find a
smaller deviation from the limiting fragmentation limit at
$\eta'<0$, although the distributions at $\eta'>0$ still keep
longitudinal scaling, where it is dominated by the quark
recombination mechanism. Figure 21 shows the comparisons of the
contributions from the gluon-gluon fusions. The results indicate the
deviation from limiting fragmentation origins from the gluon-gluon
fusion mechanism.

    Back to the $p-p$ collisions. In Fig.22 we plot our predicted curves in $p-p$ collisions from
$\sqrt s=130$GeV to $14$TeV with $\eta'$. We find that a similar
deviation from limiting fragmentation exists at $-2<\eta'<0$ if the
energies across over a big range between RHIC and LHC.

    We noted that a different deviation from limiting fragmentation at
the LHC energy is also predicted by using another kind of parton
distributions, i.e., the McLerran-Venugopalan distributions in Ref.
[32], where the same gluon-gluon fusion mechanism is used.
Therefore, investigation of a possible deviation from the limiting
curve will provide insight into the evolution equations for
high-energy QCD, although the possible larger systematic errors in
experiments may hide the deviation.

   We try to understand limiting fragmentation and its violation
from the partonic picture. The gluon distributions in Eq.(1) are
really irrelevant to the interaction energy $\sqrt s$ in the parton
model [17]. A possible relation of Eq. (1) with the interaction
energy is that the kinematic ranges of Bjorken variables $x_{1/2}$
are $\sqrt s$-dependent. To illustrate that, we draw the kinematic
ranges of two multiplying distributions at three different energies
using Eq.(2) in Fig. 23. For example, we fix $p_t=0.5 GeV$ in Eq.(2)
and take $m_{eff}=0$, thus we have $\eta=y$. We can find that at
$y=y_{bim}$ (or $y'=y-y_{bim}=0$), an extremely small
$x_1=x_{1,small}$ (or $x_2=x_{2,small}$) always combines with a
larger $x_2=x_{2,large}$ (or $x_{1,large}$). Besides,

$$x_{1,large}=x_{2,large}=\frac{p_t}{m_N}, \eqno(16)$$
is independent of $\sqrt s$, and

$$x_{1,small}=x_{2,small}=\frac{p_tm_N}{s}. \eqno(17)$$
Thus, we have

$$F_g(x_{1,large},q_t,p_t)F_g(x_{2,small},q_t,p_t)\vert_{y'=0,\sqrt s=200GeV}$$
$$\simeq
F_g(x_{1,large},q_t,p_t)F_g(x_{2,small},q_t,p_t)\vert_{y'=0,\sqrt
s=130GeV},\eqno(18)$$ since $x_{2,small}(W=130 GeV)\simeq
x_{2,small}(\sqrt s=200GeV)$. On the other hand, the difference
between the parameters $c(\sqrt s)$ at $\sqrt s=130$ and 200GeV is
small. Consequently, we have limiting fragmentation near $y'=0$ in
the gluon-gluon fusion processes in Fig. 20.

    However, an obvious difference exists near $y'=0$ if we
compare the results at the RHIC and LHC energies, i.e.,

$$F_g(x_{1,large},q_t,p_t)F_g(x_{2,small},q_t,p_t)\vert_{y'=0,\sqrt s=200GeV}$$
$$\neq
F_g(x_{1,large},q_t,p_t)F_g(x_{2,small},q_t,p_t)\vert_{y'=0,\sqrt
s=5.5TeV},\eqno(19)$$ since $x_{2,small}(\sqrt s=5.5TeV)<<
x_{2,small}(\sqrt s=200GeV)$. Although the decreasing $c(\sqrt s)$
with increasing $\sqrt s$ almost compensates for the difference in
Eq.(19), $\sigma_{in}\sim \sqrt{s}$ and $F_g\sim\sqrt{s}$ belong to
really different dynamics; therefore, the results in Fig. 21 show a
deviation from limiting fragmentation.

    It is different from the gluon-gluon fusion mechanism; the quark
recombination model naturally satisfies limiting fragmentation. The
reason is as follows. For a given $y'=\ln
x+\ln(m_N/\sqrt{m_{\pi}^2+p^2_t})$, the kinematic ranges of Eq. (7)
are $x_1\in[x,0]$ and $x_2\in[0,x]$, which are irrelevant to the
energy $\sqrt{s}$. Therefore, the resulting rapidity distributions
have limiting fragmentation, as we have shown in Fig.20.

   (ii) The possible deformation of central rapidity plateau.

    A general picture of two components of particle production in
hadron-hadron collisions predicts two types of ranges for the
distributions of final-state particles: except for limiting
fragmentation at the fragmentation ranges, particles near
midrapidity in the center-of-mass frame were expected to form a
rapidity plateau [17]. A narrow plateau appears in our
distributions, its height (but not the width) grows with energy. We
point out that the structure of the central plateau relates to the
value of parameter $m_{eff}$ rather than the parton distributions: a
larger value of $m_{eff}$ may structure a plateau with double peaks,
while it is flattened with decreasing $m_{eff}$, and the plateau
even disappears when $m_{eff}\rightarrow 0$.

    Comparing a best fitting $m_{eff}=770$MeV in $p-p$ collisions at
$\sqrt s=200$GeV (see Fig. 2) with that of $m_{eff}=400$MeV in
$Au-Au$ collisions at the same energy (see Fig. 16), the reduction
of $m_{eff}$ is possible in ultra relativistic heavy-ion collisions.
An interesting question is: whether this is a QGP effect in the RHIC
data.

    One of the most exciting research areas in RHIC collisions is to find a
new matter, i.e., the QGP, in which quarks and gluons are no longer
confined to volumes of hadronic dimensions and hadron masses are
reduced under the QCD phase transition. One of the conditions in
forming the QGP is that a sizable fraction of the initial kinetic
energy creates many thousands of particles in a limited volume.
Therefore, we consider that the QGP is formed after gluon-gluon
collision in the central region in ultrarelativistic heavy-ion
collisions. The effective mass $m_{eff}$ in Eq. (4) will be lowered
if part of the minijets go through the QGP region, since that is
where chiral symmetry is restored. We regard this decreasing value
of $m_{eff}$ as a QGP effect partly due to the restoration of chiral
symmetry. Of course, since large errors occur with the RHIC data in
the region $\eta< 3$ [31], more accurate data are necessary.
However, if the above-mentioned QGP corrections to the RHIC data are
true, one can expect a more obvious effect in $Pb-Pb$ collisions at
LHC. In Fig.24, we plot such effects in the central plateau of
$Pb-Pb$ collisions at $\sqrt s=5.5$ TeV where the values $m_{eff}$
are reduced from 400 MeV to 0.

   (iii) How large are the nuclear shadowing effects?

    It has been observed [33] that the quark distributions in a nucleus
is depleted in the low region of $x$, and this is called nuclear
shadowing. Our parton distributions [8] include nuclear shadowing
corrections through the QCD evolution equations. To illustrate the
shadowing effects in the multiplicity productions, we use the
following scaled distributions by $<N_{coll}>=<0.5N_{part}><\nu>$ to
define the nuclear shadowing factor

$$R=\frac{\frac{1}{<N_{coll}>}{\frac{dN^I_{A-A}}{d\eta}}}{\frac{dN^I_{p-p}}{d\eta}}
=\frac{\int d^2p_t\frac{d\sigma_{A-A}^I(\eta,p_t)}{\eta dp_t^2d}}
{\int d^2p_t\frac{d\sigma_{p-p}^I(\eta,p_t)}{d\eta
dp_t^2}}.\eqno(20)$$ The results in $Au-Au$ collisions at $\sqrt
s=200$GeV (solid curve) and $\sqrt s=5.5$TeV (dashed curve) are
presented in Fig.25. The calculations are stopped at larger $\eta$,
where the quark recombination mechanism and Fermi motion effects
[30] become important. The $\eta$-dependence of the nuclear
shadowing factor $R$ presents a strong energy-dependence. We find
that the shadowing corrections to the initial unintegrated gluon
distribution cannot be negligible in any studies of the nuclear
effects.

    (iv) Two component model or single component model?

    The early two component picture of hadron-hadron collisions
assumes that a broad boost-invariant central plateau is separated by
two energy-independent fragmentation regions [17]. However, the
following experiments indicate that limiting fragmentation can be
extended to beyond fragmentation ranges, and no evidence shows a
boost-invariant central plateau [25]. It seems that a
single-component mechanism  dominates the hadron production. The KLN
[6] and Albacete [7] are such single-component models. However, the
dashed curves in Figs. 3 and 4, which are determined by the evolved
unintegrated gluon distributions, are lower than the data, and it
shows that the contributions from two components of production are
necessary.

    (v) Saturation, it comes or not?

    Saturation is a limiting form of the shadowing modified gluon
distribution. In this work, we do not consider the corrections of
saturation, although shadowing is included. We compare our
predictions with those of two saturation models. Kharzeeva, Levin,
and Nardi [6] use a single production mechanism [i.e., the component
I in Eqs. (12) and (14)] to predict the heavy-ion collisions at LHC.
Instead of the two scale unintegrated gluon distribution in Eq. (1),
KLN use a saturated integrated gluon distribution

$$xG(x,p^2_t)=\left\{
\begin{array}{ll}
\frac{\kappa}{\alpha_s(Q^2_s)}Sp_t^2(1-x)^4 &p_t<Q_s(x)\\
\frac{\kappa}{\alpha_s(Q^2_s)}SQ^2_s(x)(1-x)^4
&p_t>Q_s(x)\end{array} \right.,\eqno(21)$$ to calculate the rapidity
distributions. $S$ is the inelastic cross section for the minimum
bias multiplicity and $\kappa$ is a normalization coefficient. A
free parameter is $dN/d\eta\vert_{\eta=0}$, which contains
nonperturbative information. In the KLN model, it is fixed by the
value of $dN_{Au-Au}/d\eta\vert_{\eta=0}$ at $\sqrt s=130$ GeV and
assumption

$$<\overline{\nu}(\sqrt s)>c(\sqrt s)= const. \eqno(22)$$
Figure 26 compares our results in Fig. 17 with the KLN predictions
[6].

     Albacete [7] uses the same single production mechanism but with a different
gluon distribution, which is the solution of the BK equation in the
form of single scale distribution. Albacete assumes

$$<N_{part}(\sqrt s)><\overline{\nu}(\sqrt s)>c(\sqrt s)= const.,
\eqno(23)$$ and uses the value $dN_{Au-Au}/d\eta\vert_{\eta=0}$ at
$\sqrt s=200$GeV to fixed this constant. Figure 27 compares our
results with the Albacete predictions [7].

    Except for the contributions of quark recombination and different
values of $m_{eff}$, the differences among three models originate
from the different parton distributions, which obey different QCD
evolution dynamics. Therefore, the observations of the multiplicity
distributions are useful in identifying a true QCD evolution
dynamics in the upcoming LHC data.

    (vi) Validity of the factorization formula Eq. (1).
We noted that the works in Ref. [34] presented a covariant gauge
calculation, where the transverse momentum spectrum of the gluon is
perturbatively determined by the final-state interactions of the
gluon with the nucleons in the nucleus. In this case, the
applications of the $k_t$ factorization formula (1) is unsatisfied.
However, these works neglect the QCD evolution of gluons. Our aim is
to test the predictions of the MD-DGLAP evolution equation in the
LHC physics, where a physical gauge is used for the factorization of
the evolution kernel. The above-mentioned final-state interactions
are absent, and these effects are absorbed into the phenomenological
fragmentation functions. Therefore, the application of Eq. (1) is
reasonable in this work.

\newpage
\begin{center}
\section{Summary}
\end{center}

      We use two complementary production mechanisms: hard gluon-gluon
fusion in the central rapidity region and soft quark recombination
in the fragmentation region to study the particle multiplicity
distributions in hadron-hadron collisions at high energies. We
emphasize that a set of consistent integrated and unintegrated
parton distributions, which are well defined in a broad kinematic
range are necessary in such partonic model. For this reason, our
parton distributions proposed in Ref.[8] are used. Based on the
explanations of the present data, we predict the pseudorapidity
densities of charged particles produced in $p-p$ and central $Pb-Pb$
collisions at LHC energies. We find that the limiting fragmentation
hypothesis, which generally appears in the present data of hadron
collisions is partly violated if the observations are across a wide
range between the RHIC and LHC energies. An explanation about
limiting fragmentation and its violation in partonic picture are
given. We proposed that a possible QGP effect may deform the shape
of the central plateau. In this work, we use an entirely
phenomenological parametrization of existing data to fix the energy
dependence of normalization factors of the hadronic distributions.
This simple method may need modification after we obtain the LHC
data at midrapidity. However, once this normalization is fixed,
whole particle multiplicity distributions are completely determined
by our parton distributions. The comparisons between our model and
other partonic models are given. The differences in the predicted
distributions in various models can help us to identify the true QCD
dynamics in hadron collisions.

\noindent {\bf Acknowledgments}: This work was supported by the
National Natural Science Foundations of China 10875044.

\newpage

\vskip -4.0 truecm

\centerline{\epsfig{file=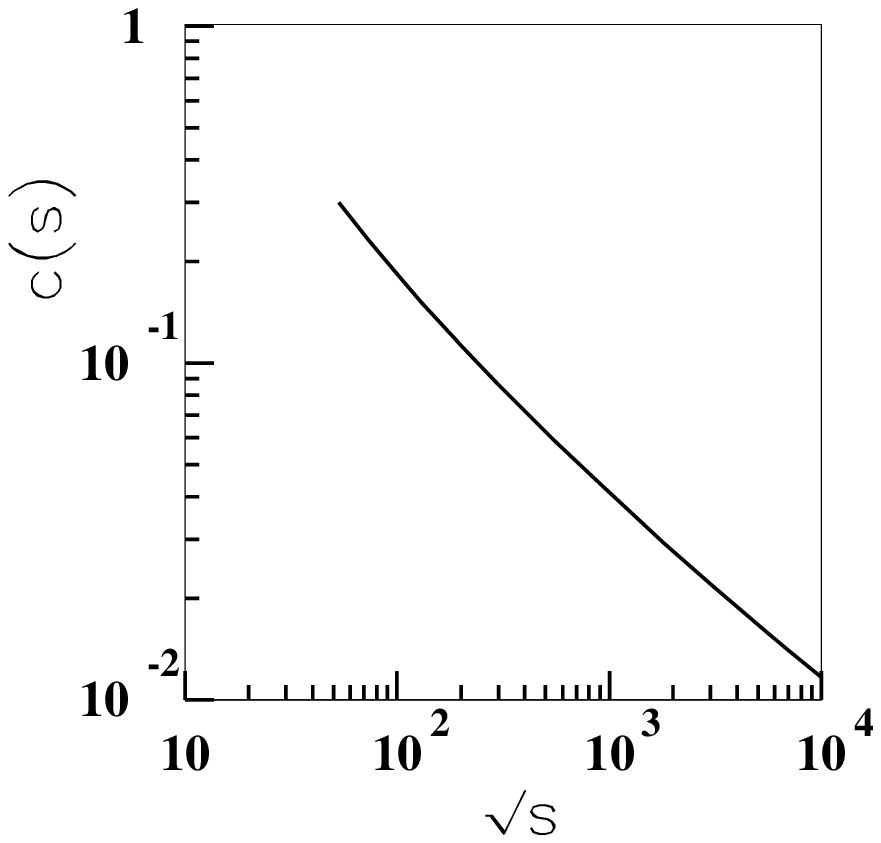,width=17.0cm,clip=}} \vskip -7.0
truecm
 \noindent
Fig. 1 Coefficient $c(\sqrt s)$ in Eq.(5) plotted as a function of
energy $\sqrt s$ using Eq.(13).

\centerline{\epsfig{file=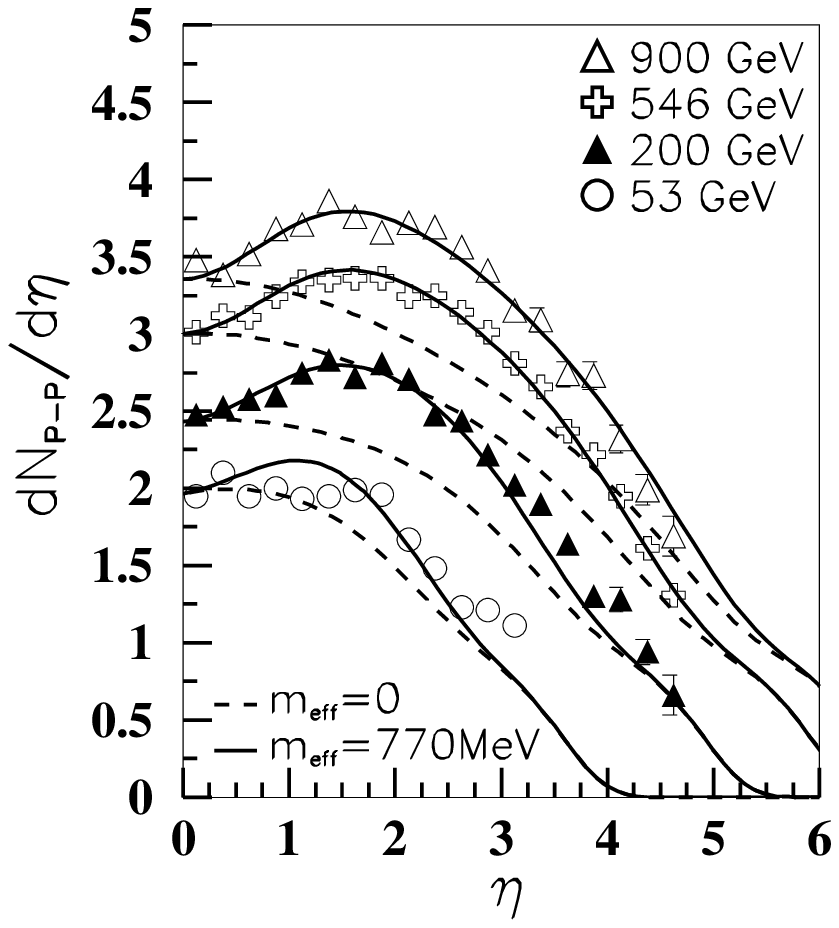,width=17.0cm,clip=}} \vskip -7.0
truecm \noindent Fig. 2 Computed pseudo-rapidity distribution of
charged particles in $p-p(\overline{p})$ collisions at various
energies with $m_{eff}=770$MeV and $\delta=0.7$ (solid curves).
The dashed curves take $m_{eff}=0$. The data are taken from Ref.
[20].

 \hbox{

\centerline{\epsfig{file=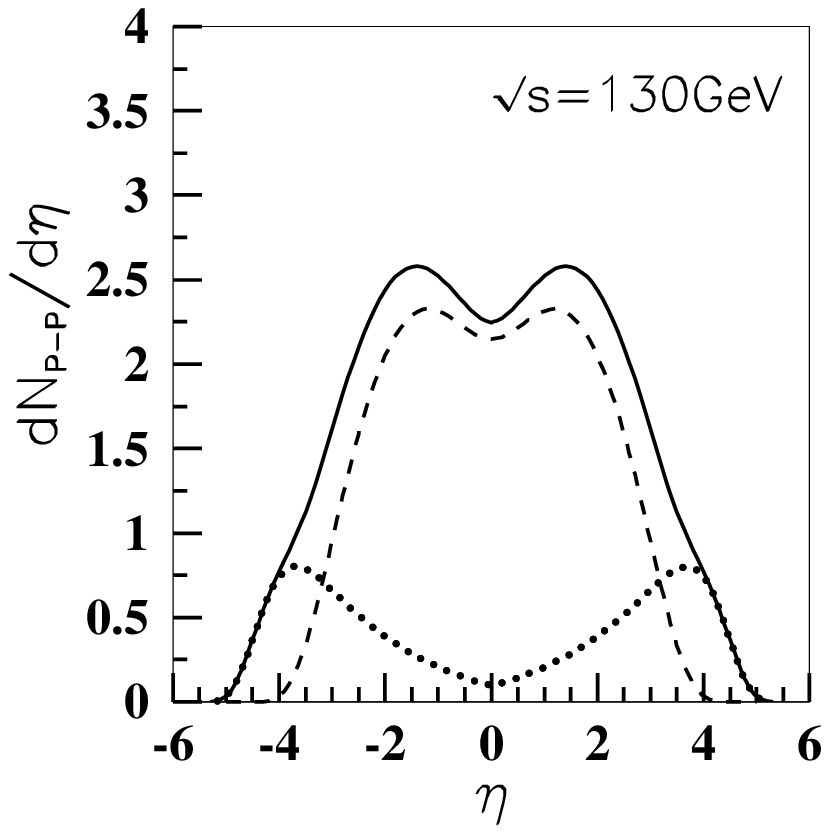,width=17.0cm,clip=}}} \vskip
-7.0 truecm\noindent Fig. 3 Solid curve in Fig.2 at $\sqrt
s=130$GeV, where dashed and dotted curves are the contributions of
gluon fusion mechanism I and quark recombination mechanism II.

 \hbox{

\centerline{\epsfig{file=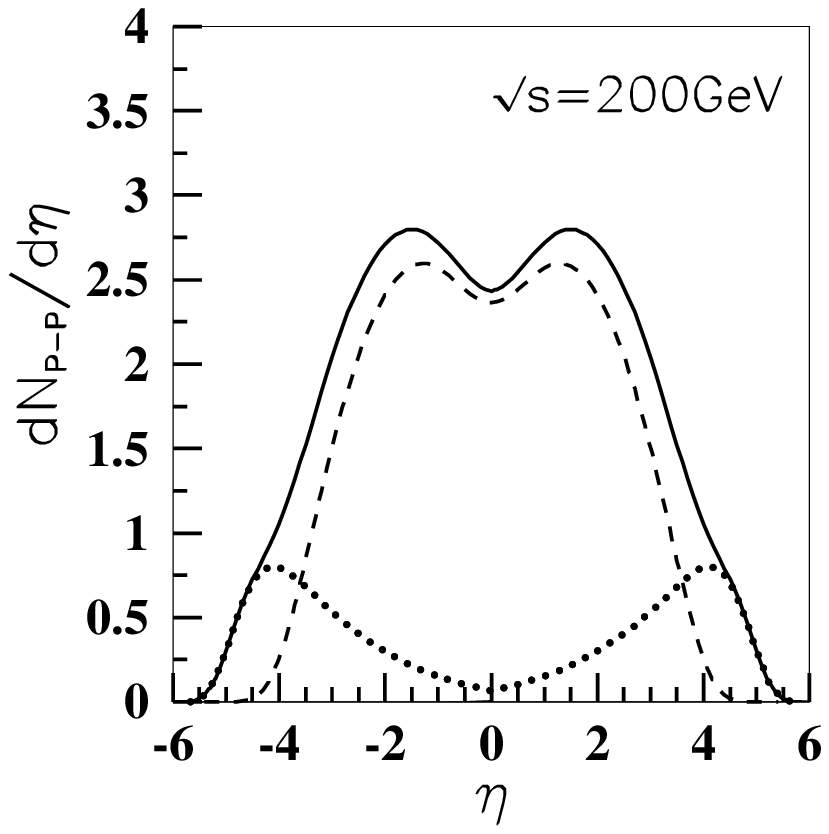,width=17.0cm,clip=}}}\vskip -7.0
truecm \noindent Fig. 4 Similar to Fig.3 but at $\sqrt s=200$GeV.

 \hbox{

\centerline{\epsfig{file=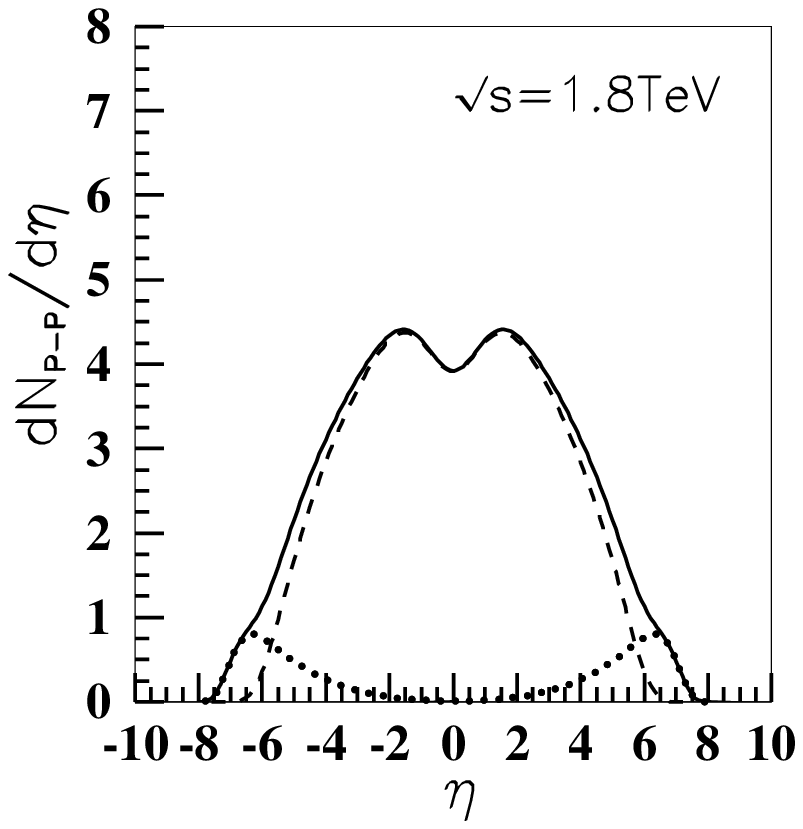,width=17.0cm,clip=}}} \vskip
-7.0 truecm\noindent Fig. 5 Predicted pseudorapidity distribution
of charged particles in $p-p$ collisions at $\sqrt s=1.8$TeV.
Dashed and dotted curves are the contributions of the gluon fusion
model (I) and quark recombination model (II).

 \hbox{

\centerline{\epsfig{file=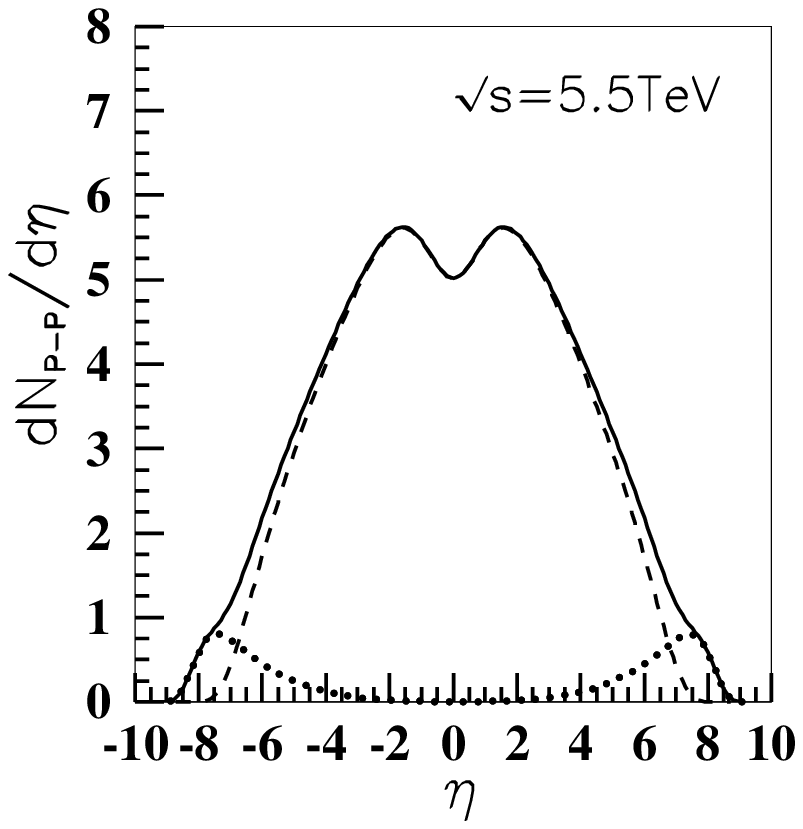,width=17.0cm,clip=}}} \vskip
-7.0 truecm\noindent Fig. 6 Similar to Fig. 5 but at $\sqrt
s=5.5$TeV.

 \hbox{

\centerline{\epsfig{file=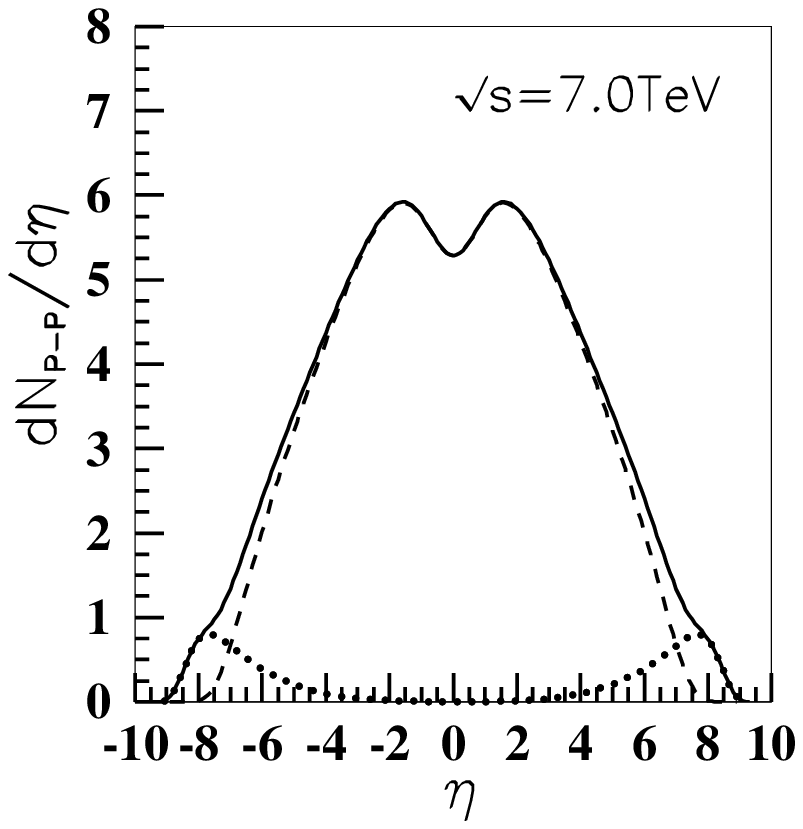,width=17.0cm,clip=}}} \vskip
-7.0 truecm\noindent Fig. 7 Similar to Fig. 5 but at $\sqrt
s=7$TeV.

 \hbox{

\centerline{\epsfig{file=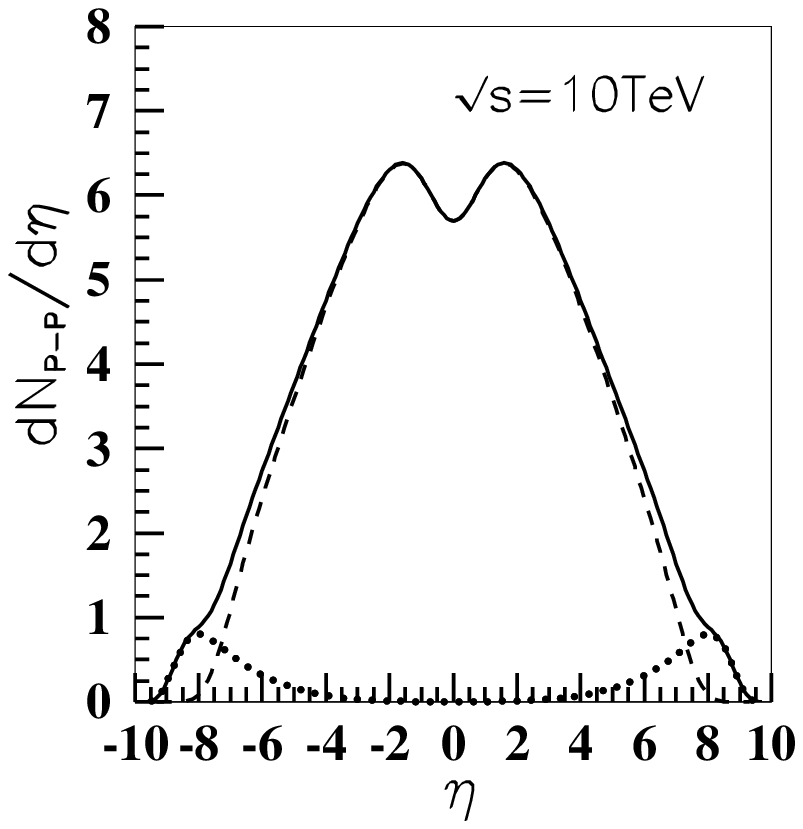,width=17.0cm,clip=}}} \vskip
-7.0 truecm\noindent Fig. 8 Similar to Fig. 5 but at $\sqrt
s=10$TeV.

 \hbox{

\centerline{\epsfig{file=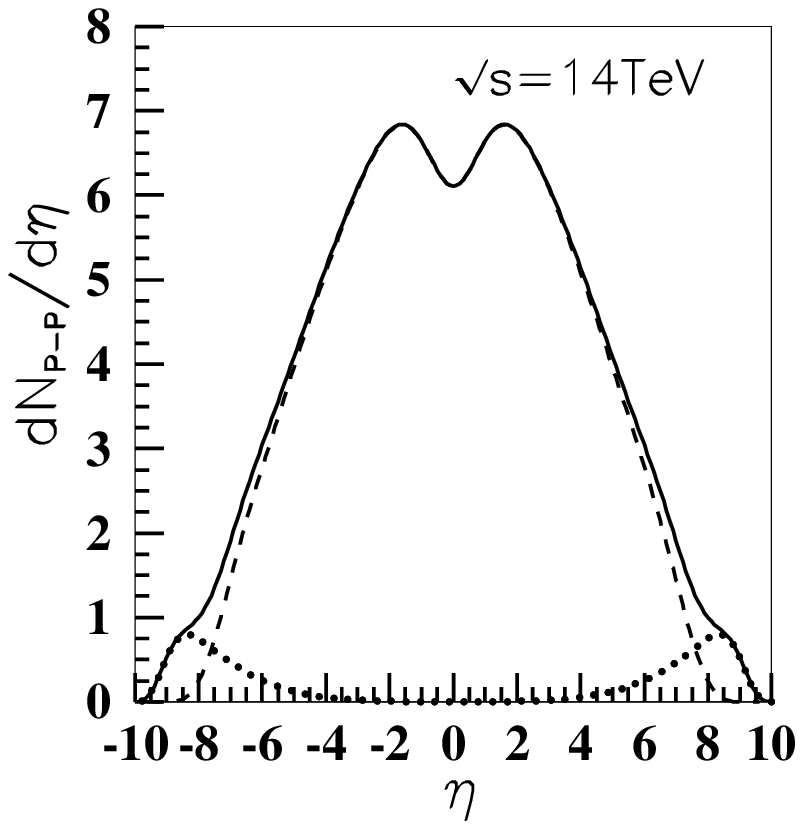,width=17.0cm,clip=}}} \vskip
-7.0 truecm\noindent Fig. 9 Similar to Fig. 5 but at $\sqrt
s=14$TeV.

 \hbox{

\centerline{\epsfig{file=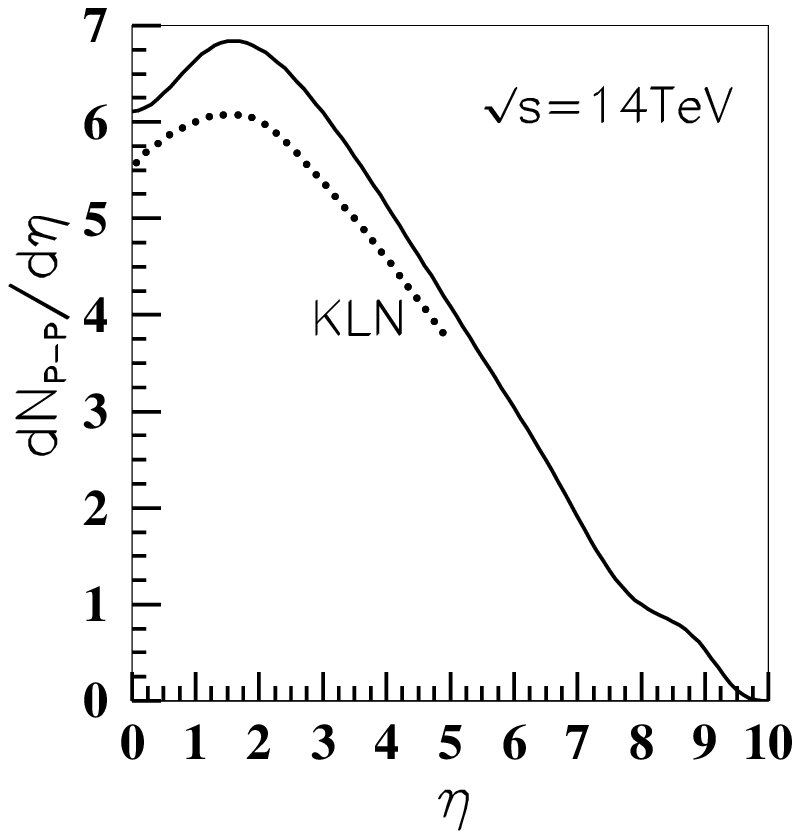,width=17.0cm,clip=}}} \vskip
-7.0 truecm\noindent Fig. 10 Comparisons of our predictions (solid
curve) for pseudorapidity distribution of charged particles in
$p-p$ collisions at $\sqrt s=14$TeV with that of KLN predictions
(dashed curve) [6].

 \hbox{
\centerline{\epsfig{file=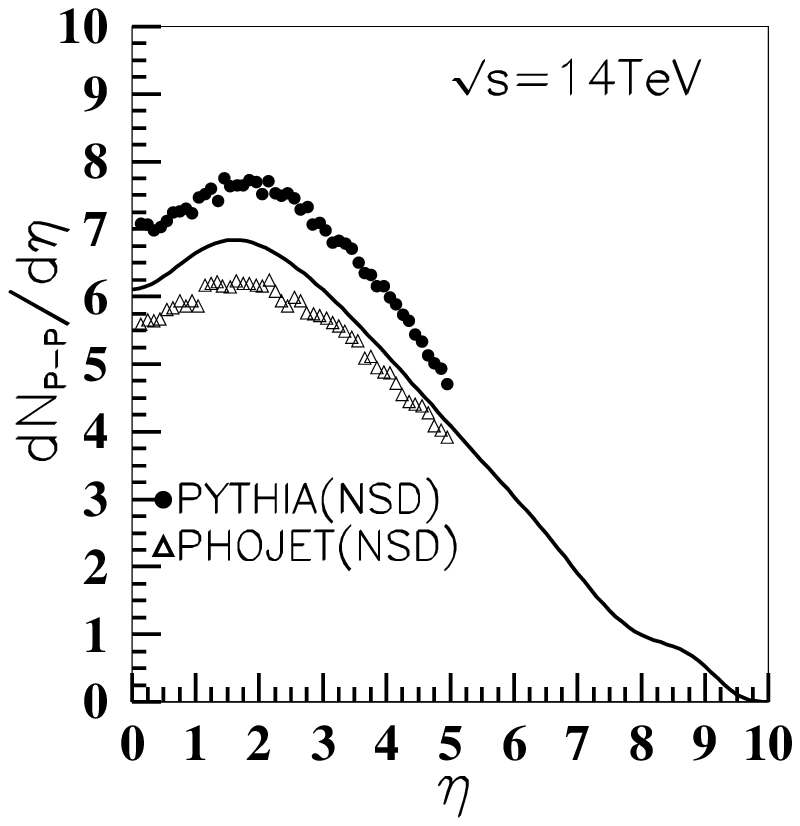,width=17.0cm,clip=}}} \vskip
-7.0 truecm\noindent Fig. 11 Comparisons of our predictions (solid
curve) for pseudo-rapidity distribution of charged particles in
$p-p$ collisions at $\sqrt s=14$TeV with that of PYTHIA [22] and
PHOJET models [23].

 \hbox{

\centerline{\epsfig{file=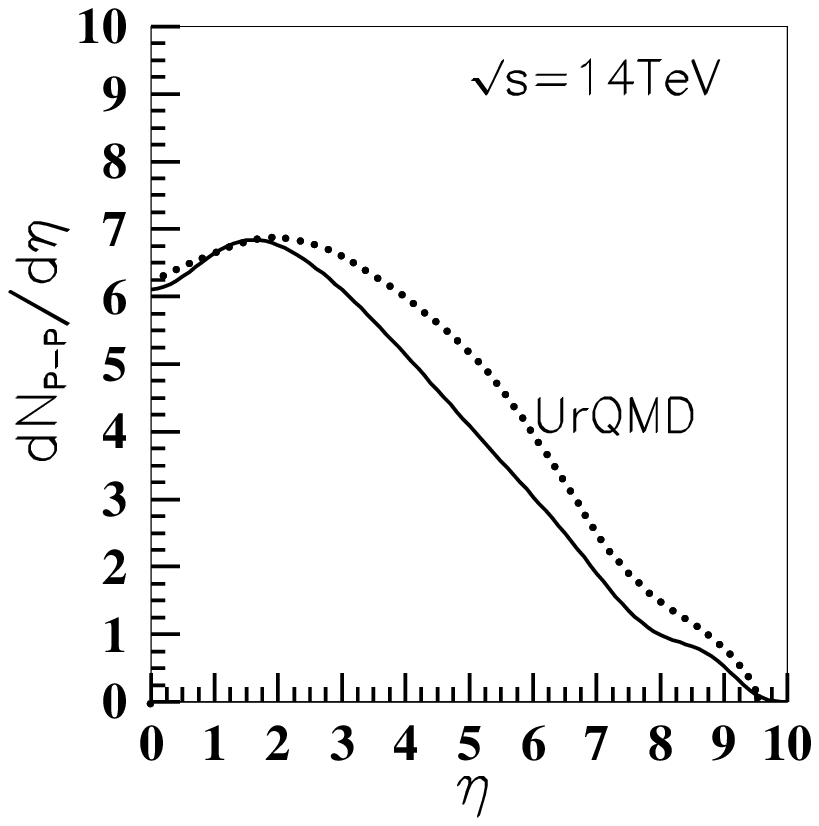,width=17.0cm,clip=}}} \vskip
-7.0 truecm\noindent Fig. 12 Comparisons of our predictions (solid
curve) for pseudorapidity distribution of charged particles in
$p-p$ collisions at $\sqrt s=14$TeV with that of the UrQMD model
(dashed curve) [24].

 \hbox{

\centerline{\epsfig{file=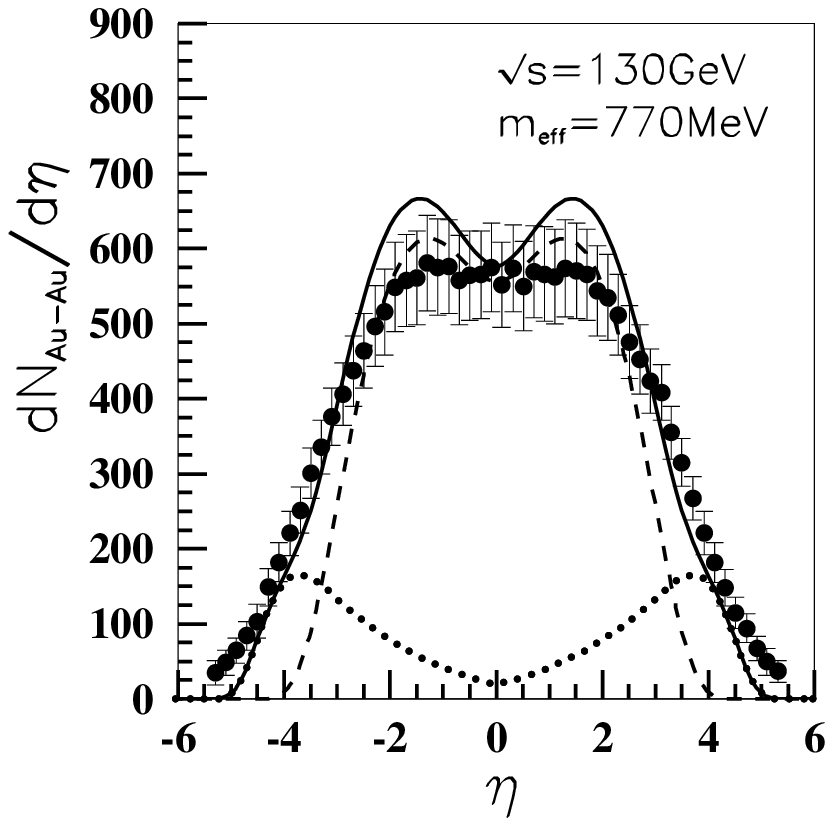,width=17.0cm,clip=}}}\vskip
-7.0 truecm \noindent Fig. 13 Pseudo-rapidity density of charged
particles produced in $Au-Au$ collisions with $0-6\%$ central cut
at $\sqrt s=130$ GeV. Data are taken from [29]. The solid curve
corresponds to $m_{eff}=770$MeV. Dashed and dotted curves are the
contributions of the gluon-gluon fusion and quark recombination,
respectively.

 \hbox{

\centerline{\epsfig{file=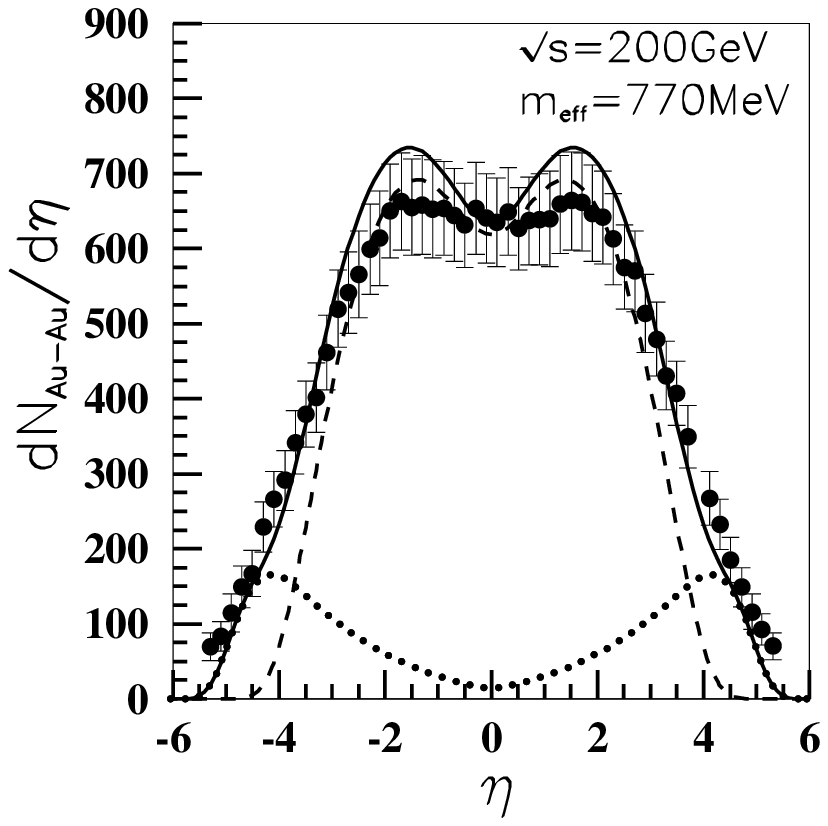,width=17.0cm,clip=}}} \vskip
-7.0 truecm\noindent Fig. 14 Same as Fig.13 but for $\sqrt
s=200$GeV.

 \hbox{

\centerline{\epsfig{file=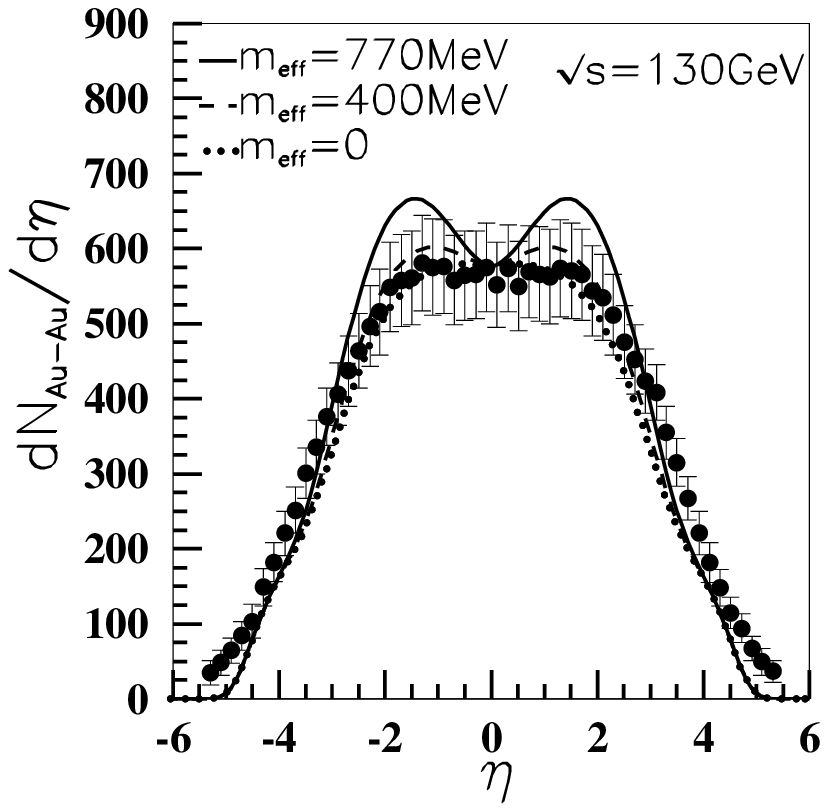,width=17.0cm,clip=}}} \vskip
-7.0 truecm\noindent Fig. 15 Pseudorapidity density of charged
particles produced in $Au-Au$ collisions with $0-6\%$ central cut
at $\sqrt s=130 GeV$ with different parameters $m_{eff}=770$MeV
(solid curve), $400$MeV (dashed curve), and $0$ (dotted curve).
Data with errors are taken from [29].

 \hbox{

\centerline{\epsfig{file=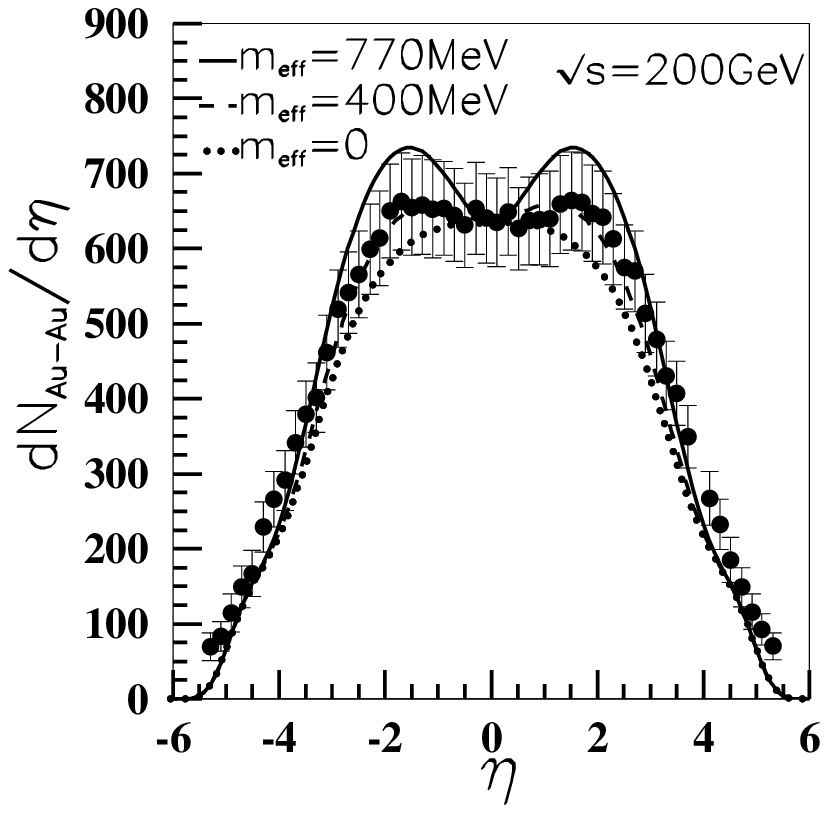,width=17.0cm,clip=}}} \vskip
-7.0 truecm\noindent Fig. 16 Same as Fig.15 but at $\sqrt
s=200$GeV.

 \hbox{
\centerline{\epsfig{file=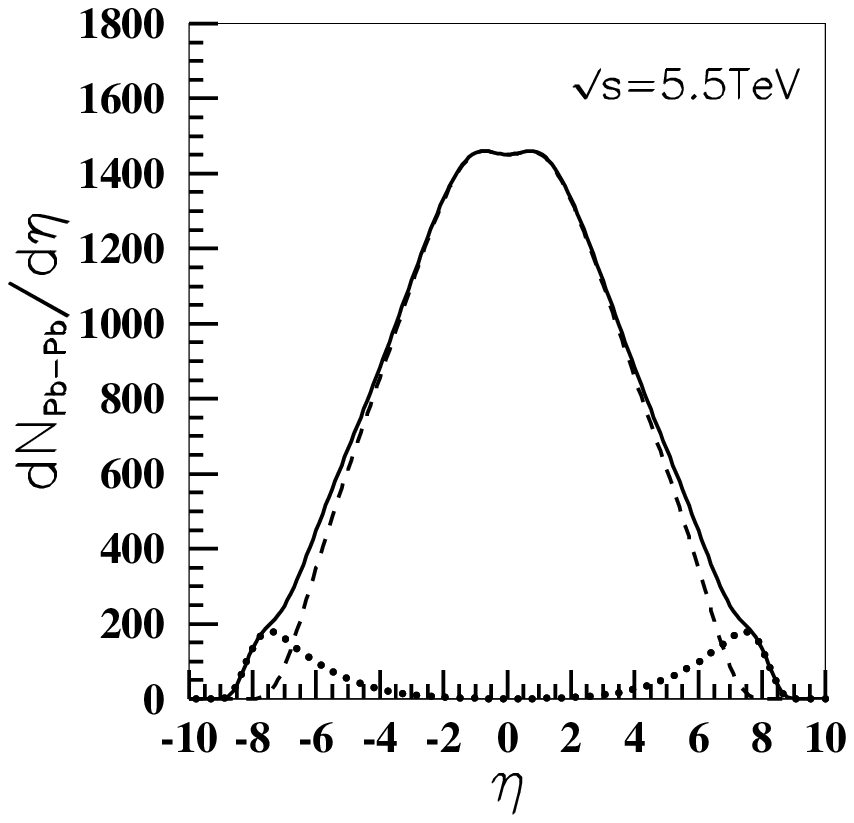,width=17.0cm,clip=}}} \vskip
-7.0 truecm\noindent Fig. 17 Predictions of pseudo-rapidity
density of charged particles produced in $Pb-Pb$ collisions with
$0-6\%$ central cut at $\sqrt s=5.5$ TeV. Dashed and dotted curves
are the contributions of the gluon-gluon fusion and quark
recombination, respectively

 \hbox{

\centerline{\epsfig{file=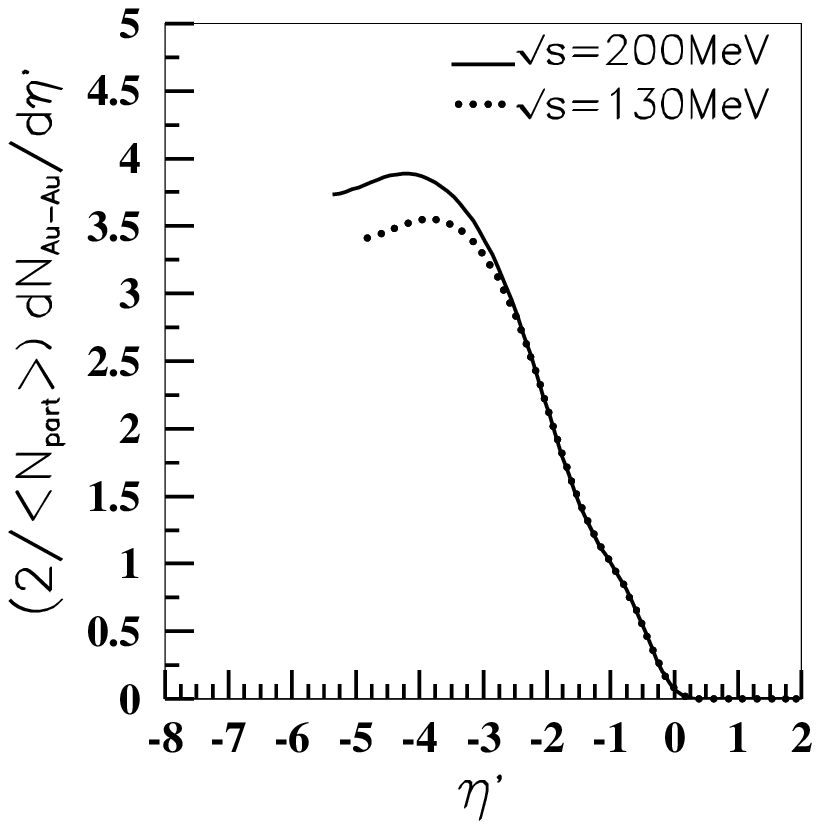,width=17.0cm,clip=}}}\vskip
-7.0 truecm \noindent Fig. 18 Shifted and scaled pseudorapidity
distribution of charged particles produced in $Au-Au$ collisions
with $0-6\%$ central cut at $\sqrt s=130$GeV (dashed curve) and
$200$GeV (solid curve). Results exhibit limiting fragmentation at
$\eta'>-1.5$.

\hbox{ \centerline{\epsfig{file=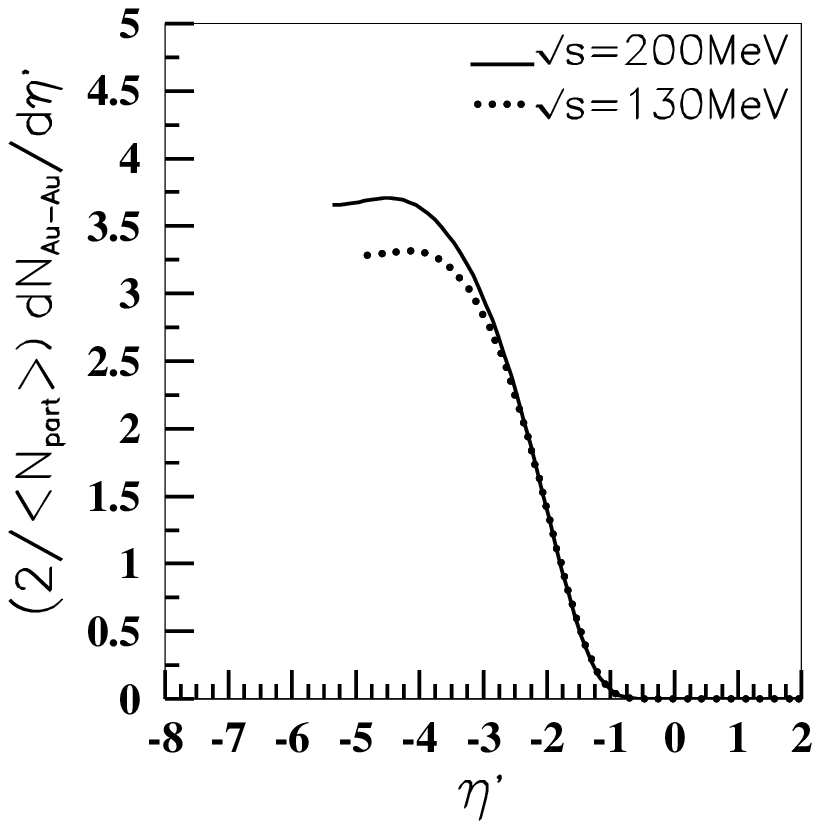,width=17.0cm,clip=}}}
\vskip -7.0 truecm\noindent Fig. 19 Similar to Fig.18 but with
contributions from gluon-gluon fusion mechanism. The same limiting
fragmentation appears.

\hbox{ \centerline{\epsfig{file=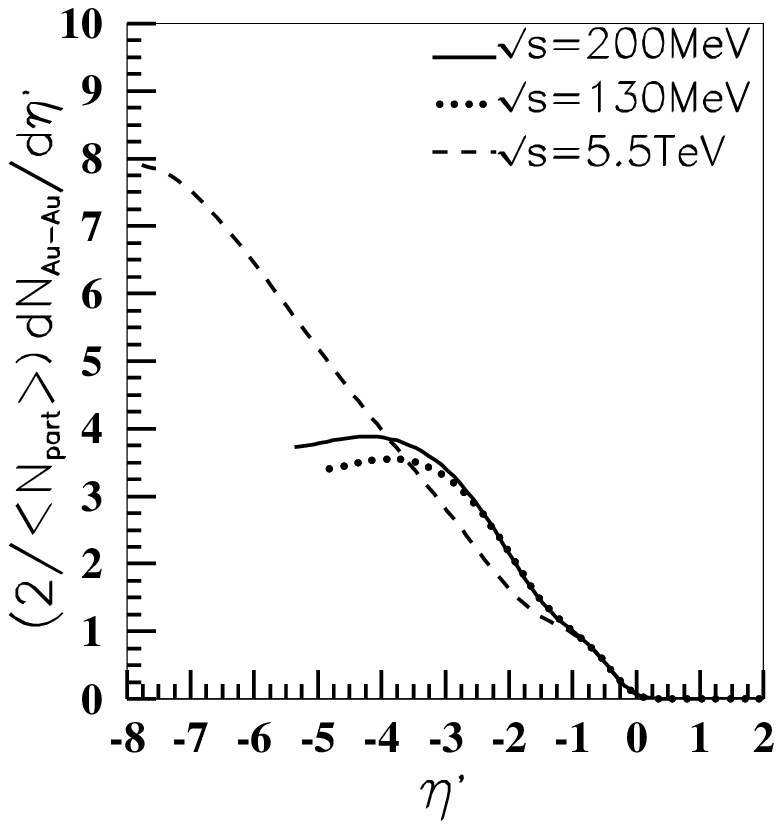,width=17.0cm,clip=}}}
\vskip -7.0 truecm \noindent Fig. 20 Comparisons of shifted and
scaled pseudorapidity distribution of charged particles produced
in $Pb-Pb$ collisions with $0-6\%$ central cut at $\sqrt s=5.5$TeV
(dotted curve) with the curves of Fig. 18. Results exhibit
deviations from limiting fragmentation at $\eta'<0$

\hbox{ \centerline{\epsfig{file=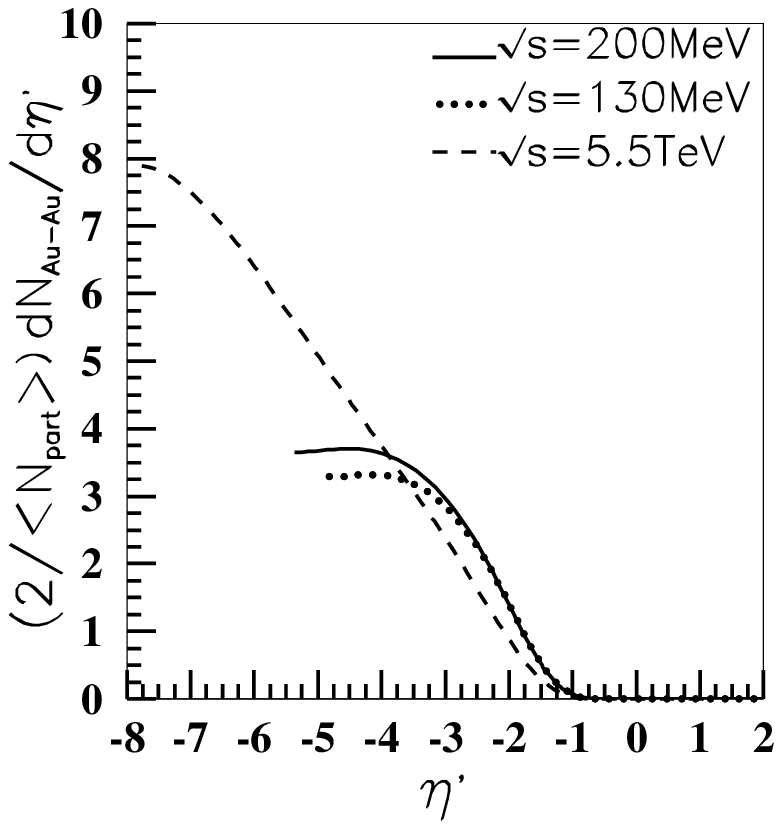,width=17.0cm,clip=}}}
\vskip -7.0 truecm \noindent Fig. 21 Similar to Fig.20 but with
contributions from the gluon-gluon fusion mechanism. Similar
violations of limiting fragmentation appear.

 \hbox{
\centerline{\epsfig{file=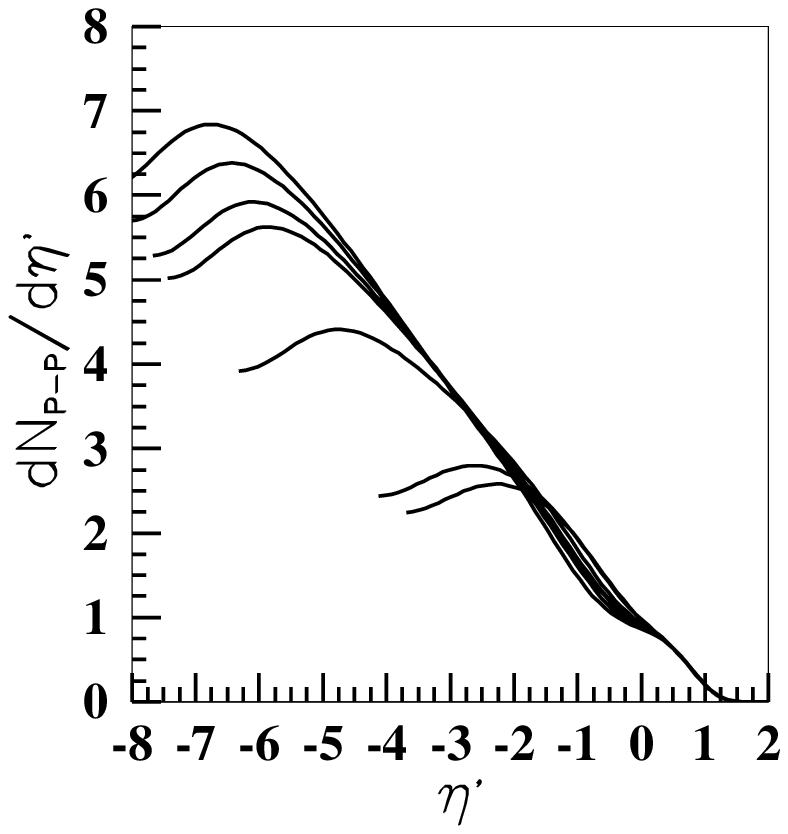,width=17.0cm,clip=}}} \vskip
-7.0 truecm \noindent Fig. 22 Predicted $dN_{N-N}/d\eta'$ at
$0-6\%$ central vs the shifted pseudo-rapidity
$\eta'=\eta-y_{beam}$ in a range of energies corresponding to
Figs.3-9 ($\sqrt s=130$GeV-14TeV from bottom to top). The results
exhibit longitudinal scaling and a small violation as in Fig. 21.

 \hbox{
\centerline{\epsfig{file=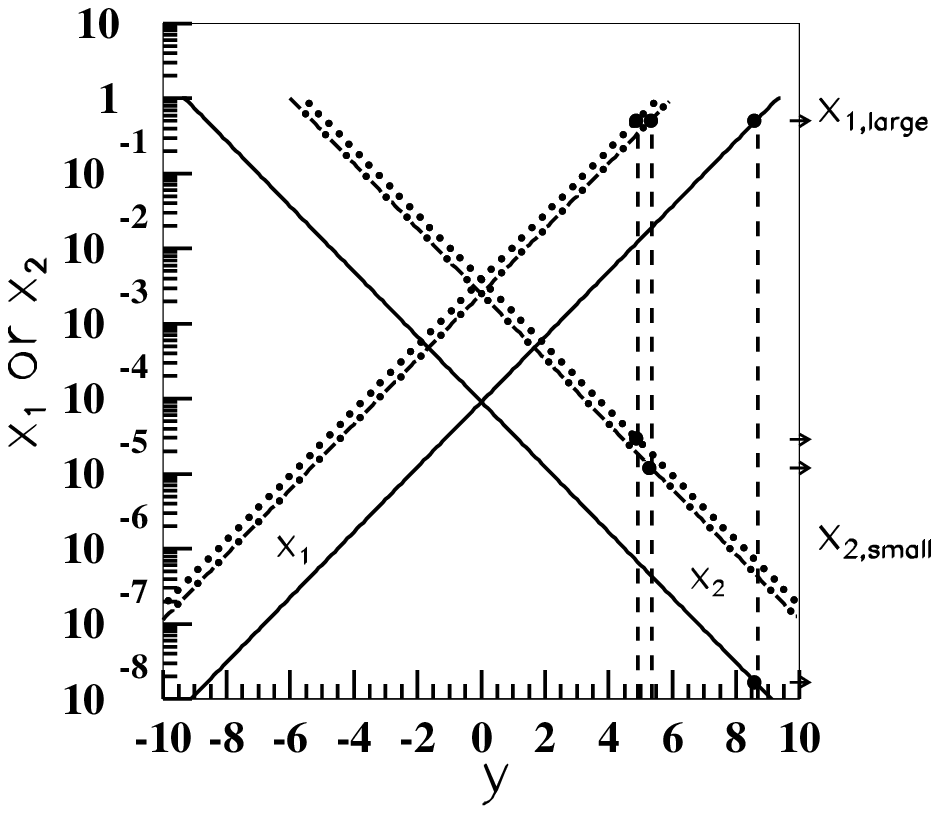,width=17.0cm,clip=}}} \vskip
-7.0 truecm \noindent Fig. 23 Kinematical ranges of $x_{1/2}$ with
different energies $\sqrt s$ and $p_t=0.5 GeV$ in Eq.(2): Solid
curves: $\sqrt s=5.5TeV$; Dashed-dotted curves: $\sqrt s=200 GeV$
and dotted curves: $\sqrt s=130 GeV$. Vertical dashed curves
correspond to (from left to right) $y'=0$ at $\sqrt s=130 GeV$,
$200 GeV$ and $5.5 GeV$.

 \hbox{
\centerline{\epsfig{file=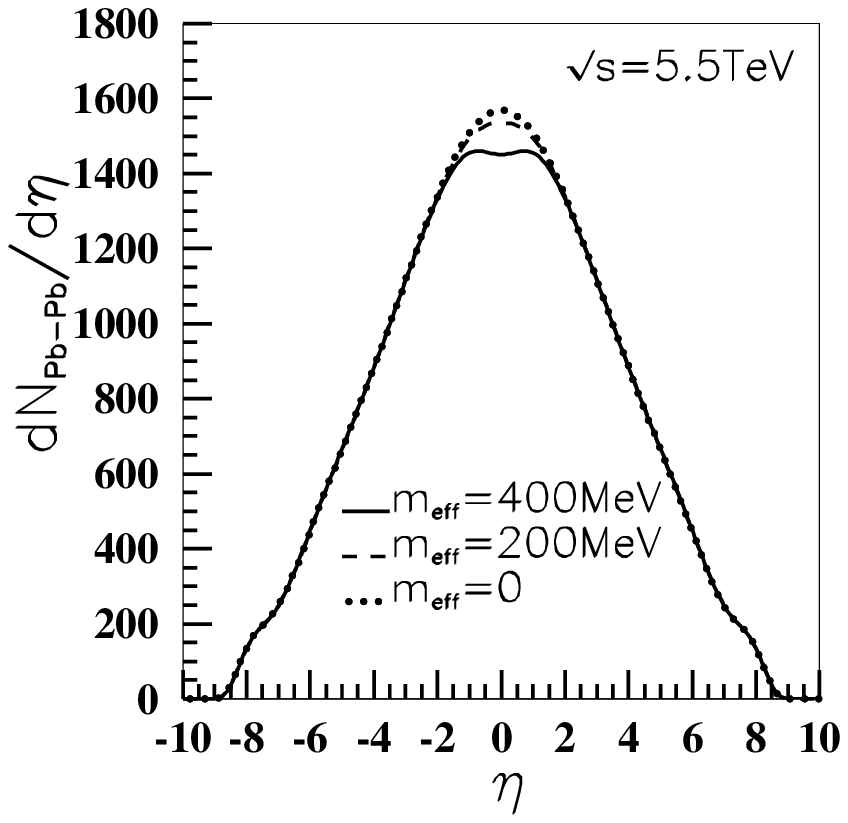,width=17.0cm,clip=}}} \vskip
-7.0 truecm
 \noindent Fig. 24 Deformation of central rapidity
plateau with decreasing parameter $m_{eff}$ in $Pb-Pb$ collisions
at $\sqrt s=5.5$ TeV. Solid, dashed, and dotted curves correspond
to $m_{eff}=400$MeV, 200MeV and $0$.

 \hbox{
\centerline{\epsfig{file=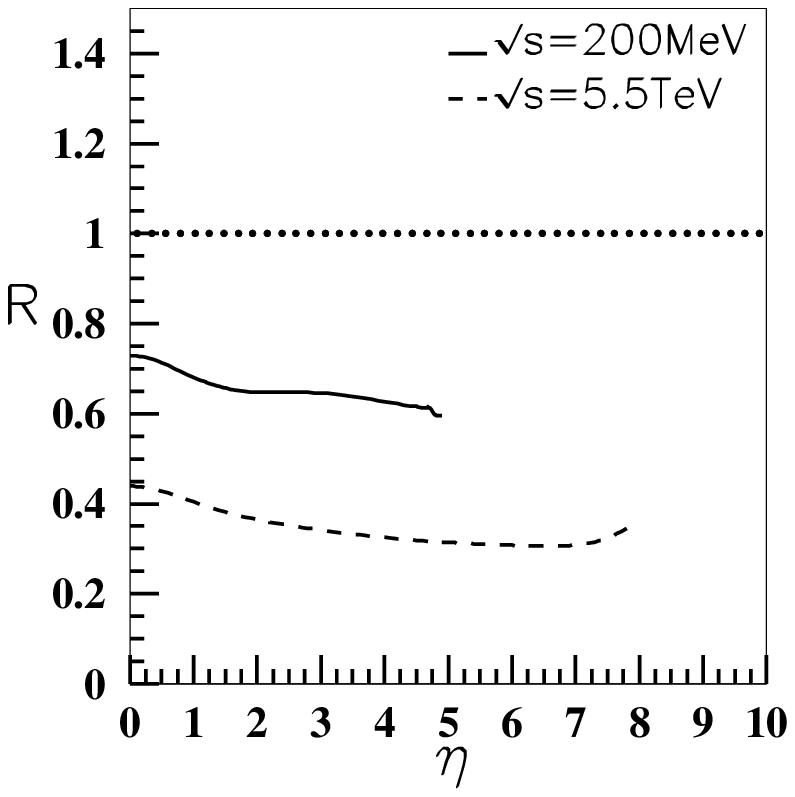,width=17.0cm,clip=}}} \vskip
-7.0 truecm
 \noindent Fig. 25 Nuclear shadowing factor $R$ in
Eq.(20) at different energies $\sqrt s$, where the pseudorapidity
distributions in $Au-Au$ collisions are scaled by $N_{coll}$.

 \hbox{
\centerline{\epsfig{file=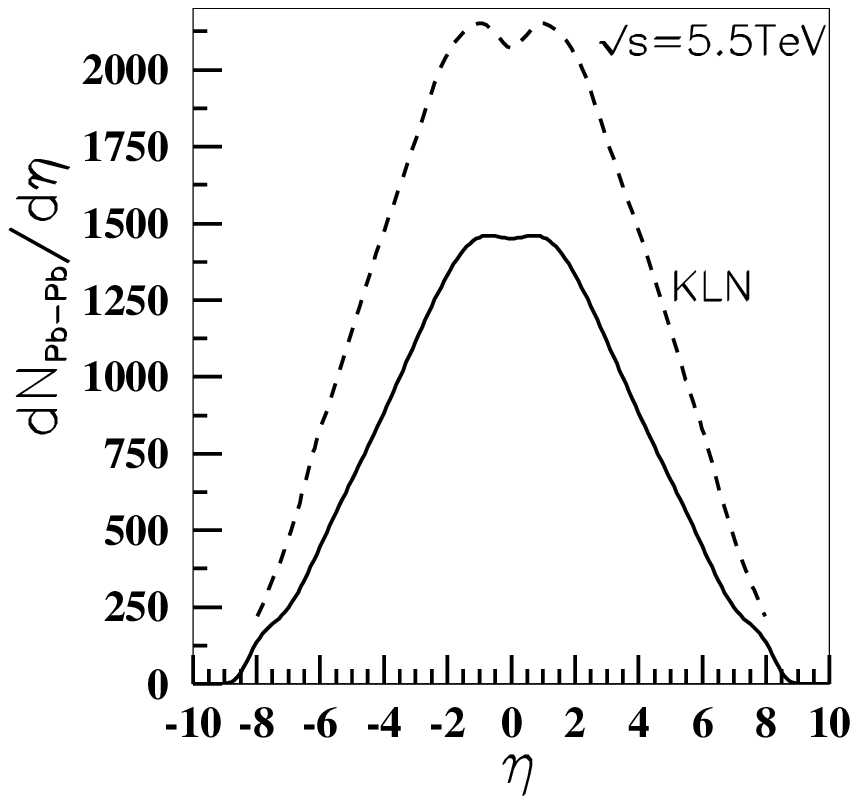,width=17.0cm,clip=}}} \vskip
-7.0 truecm
 \noindent Fig. 26 Comparison of our predictions in
Fig.17 with the results of the KLN model [6] (dotted curve), where
$m_{eff}=500$ MeV in the KLN model.

 \hbox{
\centerline{\epsfig{file=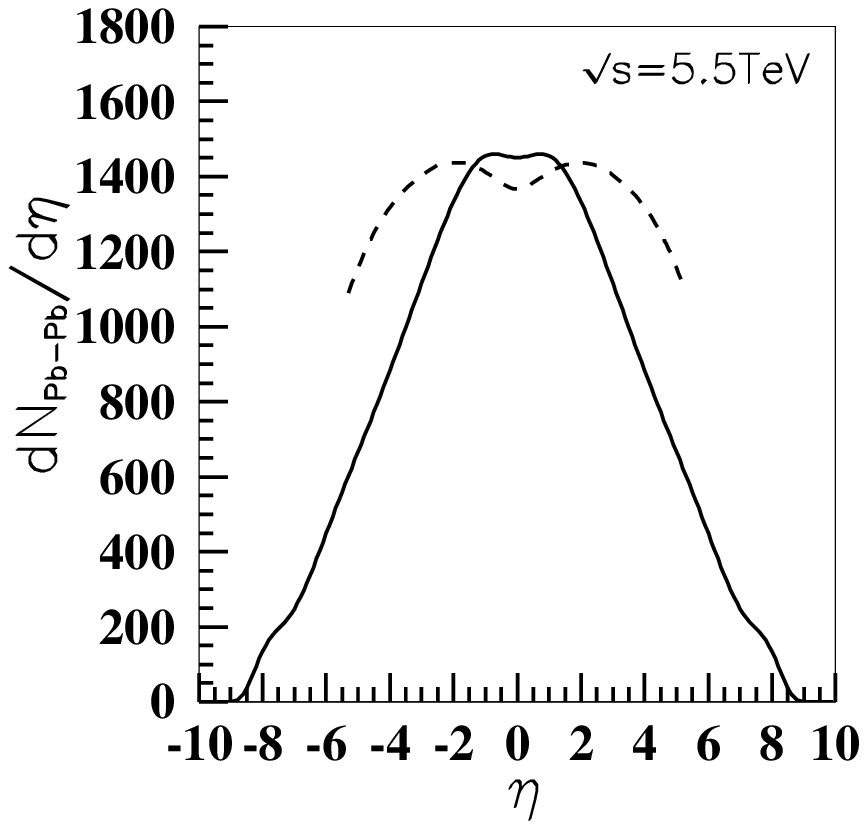,width=17.0cm,clip=}}} \vskip
-7.0 truecm

\noindent Fig. 27 Comparison of our predictions in Fig.17 with the
results of the Albacete model [7] (dotted curve), where
$m_{eff}=250 MeV$ in the Albacete model.

\end{document}